\documentclass{jetpl}
\usepackage{epsfig,amssymb,amsfonts,bm}
\twocolumn

\DeclareMathSymbol{\varSigma}{\mathord}{letters}{"06}
\newcommand{\be}{\begin{equation}}
\newcommand{\ee}{\end{equation}}

\newcommand{\vp}{\varphi}
\newcommand{\x}{{\bm x}}

\newcommand{\p}{{\bm p}}
\newcommand{\bea}{\begin{eqnarray}}
\newcommand{\eea}{\end{eqnarray}}

\title{Real-time approach to quark confined systems at finite temperatures}
\rtitle{Real-time approach to quark confined systems at finite temperatures}

\author{A.\,V.\,Nefediev}
\address{Institute of Theoretical and Experimental Physics,
117218, B.Cheremushkinskaya 25, Moscow, Russia}
\author{J.\,E.\,F.\,T.\,Ribeiro}
\address{Centro de F\'\i sica das Interac\c c\~oes
Fundamentais (CFIF), Departamento de F\'\i sica, Instituto Superior
T\'ecnico, Av. Rovisco Pais, 1049-001,
Universidade T\'ecnica de Lisboa, Lisbon, Portugal}
\rauthor{A.\,V.\,Nefediev, J.\,E.\,F.\,T.\,Ribeiro}

\abstract{A real-time formalism is proposed to incorporate finite temperatures 
into confined quark systems at the example of the Generalised
Nambu--Jona-Lasinio model for QCD. This approach
allows one to study various properties of the system at $T>0$, such as
chiral symmetry breaking and restoration, properties of the bound-state
spectrum, and so on.}


\begin{document}
\maketitle

\section{Introduction}

While the property of colour confinement is expected from general
considerations and is supported by lattice calculations, no
effective theory exists which derives confinement directly from
the QCD Lagrangian. Therefore, in the absence of controllable
analytic solutions of QCD, one has to resort to models to describe
hadronic phenomenology. However, these models must satisfy a very
stringent set of minimal requirements. They must be: i)
relativistic, ii) chirally symmetric, and iii) able to provide a
mechanism of spontaneous breaking of chiral symmetry, iv) they
should contain confinement, in order to be able to address the
issue of excited states. To satisfy all
the above requirements with one and the same model, turns out to be
a highly nontrivial matter. 
There is a model, however, which does incorporate the
aforementioned set of requirements. This is the
Generalised Nambu--Jona-Lasinio (GNJL) model with the
instantaneous vector confining kernel
\cite{Orsay,Lisbon} and confinement of quarks guaranteed due to an
instantaneous infinitely rising (for example, linear) potential,
to which extra ingredients can be added like, for instance, the colour
Coulomb potential. Then chiral symmetry breaking can
be described by standard summation of valence quark
self-interaction loops (the mass--gap equation), while hadrons are
obtained from the Bethe--Salpeter equation for the
multiquark bound states. Besides, GNJL is known to fulfill
the well--known low--energy theorems of Gell-Mann, Oakes, and
Renner, Goldberger and Treiman, Adler
selfconsistency zero, the Weinberg theorem,
and so on --- see Ref.~\cite{lowen}. For highly excited
hadrons this model predicts effective restoration of chiral symmetry
\cite{parity2} in accordance with general expectations \cite{G6}. At this
stage, we would like to emphasise the universal nature of the above
low--energy theorems as well as of the chiral restoration in excited
hadrons regardless of the actual gluonic interactions
that would eventually be responsible for chiral symmetry breaking. Thus, 
the GNJL model is expected to be quite useful to describe all those
phenomena which are essentially driven by the same global chiral
symmetry.

Incorporation of finite chemical potentials and finite temperatures stand as
important forward steps in the development of this class of models which 
not only help us into solving various puzzles posed by experimental 
results but, as importantly, they may be of assistance in predicting new
phenomena for experimental testing. A number of attempts has been undertaken
to include finite quark densities into consideration \cite{GW1,Ad1}.
By contrast, the incorporation of finite temperature appears to be a slightly
more involved problem. Indeed, one may encounter technical problems applying
techniques such as the Matsubara finite-temperature approach to the models,
where
only real-time calculations are possible. Thus we adhere to a
different approach, based on the formalism of Bogoliubov--Valatin
transformations, which is a natural mathematical tool to deal with GNJL systems.
Such an approach to chiral
symmetry breaking at $T=0$ was
successfully used in a series of papers \cite{Lisbon} and then extended
to redefine the theory entirely in terms of the
quark--antiquark bound states \cite{replica4}.
The underlying idea of the method is to consider the true thermal vacuum of the
system, as a collective (coherent-like) phenomenon of zero-temperature chirally
symmetric theory eigenstates. To this end it is necessary to build a
pseudounitary, temperature-dependent, transformation relating the thermal
vacuum with the
zero-temperature unbroken vacuum (such a real-time formalism for not confined
systems
was suggested in Ref.~\cite{TakUm}). For quarks, this transformation leads to
the formation
of new effective objects --- dressed quarks which, when seen from the
zero-temperature chirally symmetric
Fock space, look like original bare quarks ``escorted by'' an
accompanying cloud of quark--antiquark pairs. 
The cornerstone of this method, at $T=0$, amounts to finding, through a
variational calculus, the minimal vacuum energy of the theory evaluated as a
functional of the order parameter, called the chiral angle. Then in order to
include finite temperature effects we have to consider the free energy $F_{\rm
vac}=E_{\rm
vac}-TS$, with $S$ being the entropy of the vacuum, instead of the usual $E_{\rm
vac}=\langle 0|H|0\rangle$. For a given
temperature $T$ this minimisation process ensures a proper balance between the
vacuum energy and the entropy of the vacuum, so that the resulting vacuum state
is stable, with $F_{\rm vac}<F_{\rm vac}^{(0)}$, where $F_{\rm vac}^{(0)}$ is
the free energy of the trivial, unbroken, vacuum.

\section{GNJL model at $T=0$}\label{sec2}

We start with a short introduction to the GNJL chiral quark model
\cite{Orsay,Lisbon} given by the Hamiltonian 
\begin{eqnarray}
H&=&\int d^3 x\bar{\psi}({\bm x},t)\left(-i{\bm \gamma}\cdot
{\bm \bigtriangledown}+m\right)\psi({\bm x},t)\nonumber\\
&+&\frac12\int
d^3 xd^3y\;J^a_\mu({\bm x},t)K_{\mu\nu}({\bm x}-{\bm y})J^a_\nu({\bm y},t),
\label{H} 
\end{eqnarray}
where
$J_{\mu}^a({\bm x},t)=\bar{\psi}({\bm
x},t)\gamma_\mu\frac{\lambda^a}{2}\psi({\bm x},t)$.
Having in mind that the qualitative results and conclusions do
not depend on any particular form of the quark kernel
$K_{\mu\nu}({\bm x}-{\bm y})$, we only require it to be confining
and to introduce a natural scale. For
phenomenological applications this scale is to be fixed of order
of $300\div 400$ MeV. No extra constraints are imposed on the
kernel. 

Among possible quark--quark interactions described by the
Hamiltonian (\ref{H}) we have quark selfinteractions.
It turns out that these self-interactions are removed by the use of an 
appropriate Bogoliubov--Valatin transformation from bare to dressed quarks,
which can be conveniently parametrised by means of the so-called chiral angle
$\vp_p$ ($p$ being
the relative momentum in the dressing quark--antiquark
pairs) \cite{Orsay,Lisbon}:
$$
\psi^\alpha(\x)=\sum_{p,s}e^{i\p\x}
[b^\alpha_{ps}u_s(\p)+d^{\alpha\dagger}_{ps} v_s(-\p)],
\label{psi}
$$
\begin{eqnarray}
u(\p)&=&\frac{1}{\sqrt{2}}\left[\sqrt{1+\sin\vp_p}+
({\bm\alpha}\hat{\p})\sqrt{1-\sin\vp_p}\right]u_0(\p),\nonumber\\
v(-\p)&=&\frac{1}{\sqrt{2}}\left[\sqrt{1+\sin\vp_p}-
(\bm{\alpha}\hat{\p})\sqrt{1-\sin\vp_p}\right]v_0(-\p)\nonumber.
\end{eqnarray}
where $\alpha$ is the colour index of $N_C$ colours. It is
convenient to define the chiral angle varying in the range
$-\pi/2<\vp_p\leqslant\pi/2$ with the boundary
conditions 
\be
\vp_p(p=0)=\pi/2,\quad \vp_p(p\to\infty)\to 0. 
\label{bc}
\ee

The normal ordered Hamiltonian (\ref{H}) becomes:
\be
H=E_{\rm vac}+:H_2:+:H_4:,
\label{H3}
\ee
and the usual procedure is to demand the quadratic part
$:H_2:$ to be diagonal or, equivalently, that the vacuum
energy $E_{\rm vac}$ should become a minimum. Then the
corresponding mass-gap equation,
\be
\delta E_{\rm vac}[\vp]/\delta\vp_p=0,\quad 
E_{\rm vac}[\vp]=\langle 0[\vp]|H|0[\vp]\rangle,
\label{mge}
\ee
ensures the anomalous Bogoliubov
terms $b^\dagger d^\dagger$ and $db$ to be absent in
$:H_2:$.
As soon as the mass-gap equation is solved and a nontrivial chiral
angle $\tilde{\varphi}_p$ is found, the Hamiltonian (\ref{H3}) takes a diagonal
form,
\be
H=E_{\rm vac}+\sum_{p,\alpha,s}
E_p[\tilde{b}^{\alpha\dagger}_{ps}\tilde{b}^\alpha_{ps}+
\tilde{d}^{\alpha\dagger}_{ps}\tilde{d}^\alpha_{ps}],
\label{H2diag}
\ee
where $E_p$ is the dressed-quark dispersive law and the quark creation and
annihilation operators
$\tilde{b}^{\alpha\dagger}_{ps}$, $\tilde{b}^\alpha_{ps}$,
$\tilde{d}^{\alpha\dagger}_{ps}$, and $\tilde{d}^\alpha_{ps}$ are
consistent with
the Fock space $\tilde{\cal{F}}$ built on top of the
nontrivial BCS vacuum $\tilde{|0\rangle}\equiv |0[\tilde{\varphi}]\rangle$. The
contribution of the
$:H_4:$ part is suppressed as
$1/\sqrt{N_C}$, so this completes the diagonalisation of the Hamiltonian
(\ref{H}) in the
quark sector (at the BCS level).

Since the definition of a one-particle state has changed dramatically, the  
vacuum $|\tilde{0}\rangle$ annihilated by the dres\-sed single-particle
operators is also
different from the trivial vacuum $|0\rangle_0\equiv |0[\vp=0]\rangle$.
Indeed, as it always happens after a Bogoliubov--Valatin transformation, the
true vacuum, with the minimal vacuum energy, contains an infinite set of
strongly correlated
$^3P_0$ quark-antiquark pairs \cite{Lisbon},
\be
|\tilde{0}\rangle=e^{Q_0^\dagger-Q_0}|0\rangle_0,\quad
Q_0^\dagger=\frac12\sum_p\vp_pC_p^\dagger,
\label{S0}
\ee
where $C_p^\dagger=b^{\alpha\dagger}_{ps}[({\bm \sigma}\hat{{\bm
p}})i\sigma_2]_{ss'}d^{\alpha\dagger}_{ps'}$ with
$\sigma$'s being the $2\times 2$ Pauli matrices.
The operator $C_p^\dagger$ creates a $^3P_0$ quark-antiquark pair with zero
total
momentum and the relative three-dim\-ensional momentum $2{\bm p}$. The chiral
angle $\tilde{\vp}_p$ is a solution to the mass-gap equation (\ref{mge}) and it
\lq\lq measures" the weight of the pairs with the given relative momentum, so
that the operator $\exp[Q_0^\dagger-Q_0]$ creates a cloud of correlated pairs,
and the BCS
vacuum $|\tilde{0}\rangle$ has the form of a coherent-like state when seen from
the point of view of the naive Fock space ${\cal{F}}_0$ (that is, the Fock space
built on top of the trivial vacuum $|0\rangle_0$). The quark Fock space
$\tilde{\cal{F}}$ is built over $|\tilde{0}\rangle$ by
repeated application of quark/antiquark creation operators.

Using the commutation relations for the quark operators $b$ and $d$, 
one can easily find the following representation for the
new vacuum \cite{Lisbon}:
\be
|0[\varphi]\rangle=\mathop{\prod}\limits_{p}\left[\sqrt{w_{0p}}+
\frac{1}{\sqrt{2}}\sqrt{w_{1p}}C^\dagger_p
+\frac12\sqrt{w_{2p}}C^{\dagger 2}_p\right]|0\rangle_0,
\label{nv}
\ee
where the coefficients
\be
w_{0p}=\cos^4\frac{\vp_p}{2},\,
w_{1p}=2\sin^2\frac{\vp_p}{2}\cos^2\frac{\vp_p}{2},\,
w_{2p}=\sin^4\frac{\vp_p}{2},
\label{ws}
\ee
clearly obey the condition $w_{0p}+w_{1p}+w_{2p}=1$ and thus they admit the
following natural
interpretation: they represent the corresponding
probabilities of having in the new vacuum zero, one, and two
quark--antiquark pairs with a given relative momentum. Notice that powers of the
operator $C_p^\dagger$ higher than two cannot appear because of the Fermi
statistics for quarks. 
Then it is easy to show that the new vacuum is normalised,
\be\label{norm00} 
\langle 0{[\varphi]}|0{[\varphi]}\rangle=\prod_p(w_{0p}+w_{1p}+w_{2p})=1,
\ee
and that it is orthogonal to the trivial vacuum in the limit of infinite volume
of the space $V$:
\be
\langle
0{[\varphi]}|0\rangle_0=\exp\left[\sum_p
\ln\left(\cos^2\frac{
\vp_p}{2}\right)\right]
\mathop{\longrightarrow}\limits_{V\to \infty}0.
\ee

\section{GNJL model at finite temperatures}\label{sec3}

As it was mentioned before, consideration of finite temperatures 
amounts to evaluation of both the vacuum energy and the entropy of the system
simultaneously and then to the minimisation of the free energy with respect to
the
order parameter. With the help of Eq.~(\ref{nv}) one can easily
find for the entropy of the vacuum:
\be
S=-N_CN_f\sum_p\sum_{n=0}^2w_{np}\ln w_{np},
\label{entr}
\ee
where the probabilities $w_{np}$ are defined in Eq.~(\ref{ws}). 

Although we shall be working only with the c-number (\ref{entr}), it is
instructive to see how it appears as a result of averaging
over the vacuum of a local operator, $S=\langle 0[\vp]|K|0[\vp]\rangle$, which
should: i) be built
entirely in terms of the operators $C_p$ and $C_p^\dagger$
 --- the only building blocks with the quantum numbers of the vacuum at our
disposal; ii) commute with the Hamiltonian (\ref{H2diag}); 
iii) comply with the relation (\ref{entr}). The final
result reads:
$$
K=-\sum_p\left[\ln w_{0p}+\frac12C_p^\dagger
C_p\ln\frac{w_{1p}}{w_{0p}}
+\frac12C_p^{\dagger 2}C_p^2\ln\frac{w_{2p}}{w_{1p}}
\right].
$$

As it was mentioned before, the thermal mass-gap equation guarantees the balance
between the energy and the entropy of the vacuum and can be derived by
minimising the free energy operator $H-TK$ averaged over the vacuum
$|0[\vp]\rangle$. Then, with the help of Eqs.~(\ref{mge}) and (\ref{entr}), one
can find for the free energy:
\be
F_{\rm vac}[\vp]=\langle 0[\vp]|(H-TK)|0[\vp]\rangle=E_{\rm vac}-TS.
\label{omeg0}
\ee
Then the mass-gap equation is
\be
\delta F_{\rm vac}[\vp]/\delta\vp_p=\delta E_{\rm vac}[\vp]/\delta\vp_p-
T\delta S[\vp]/\delta\vp_p=0,
\label{mget0}
\ee
which is nothing but a generalisation of the zero-tem\-perature mass-gap
equation
(\ref{mge}) to finite temperatures. 
It is easy to find by a straightforward calculation that
\be
\delta S/\delta\vp_p=-\sin\vp_p\left[\ln\left(\tan^2\frac{\vp_p}{2}
\right)+\cos\vp_p\;\ln 2\right].
\label{dmge}
\ee

Mass-gap equation (\ref{mget0}) with the entropy and its variation given by
Eqs.~(\ref{entr}) and (\ref{dmge}) is
the central result of this paper. 
Its physical interpretation is straightforward:
${}^3P_0$ quark--antiquark pairs are condensed
in the vacuum to lower the vacuum energy and to break chiral symmetry. With 
temperature, some pairs are ``removed'' from the vacuum so to absorb the heat
that has been added to the system. The chiral condensate decreases (evaporates)
accordingly and chiral symmetry gets less broken. The mass-gap equation
guarantees the proper balance between these two opposite processes.
Concluding this discussion, let us remind the readers that the chiral
angle is nothing but the wave function of the Goldstone boson
($\Psi_\pi=\sin\vp_p$ \cite{Orsay,Lisbon}) responsible for the chiral symmetry
breaking. With the proper solution of the thermal mass-gap 
equation in hands, one is able to study microscopically the process of the
Goldstone boson ``melting'' with the 
temperature increase.

The extra terms in the mass-gap equation which stem
from the temperature are consistent with the boundary conditions (\ref{bc})
imposed on the
chiral angle. Indeed, as clearly seen from Eq.~(\ref{dmge}),
\be
\left(\delta S/\delta\vp_p\right)_{|\vp_p=\pi/2}=
\left(\delta S/\delta\vp_p\right)_{|\vp_p=0}=0.
\ee
Since equation $\delta S/\delta\vp_p=0$ does not have other
solutions, the entropy increases monotonously from its minimum at $\vp_p=0$ to
its maximum at $\vp_p=\pi/2$.
Then Eq.~(\ref{mget0}) embodies two natural limits: in the
infinite quark mass limit one must have $\vp_p =\pi/2$ for all $p$'s, and
indeed we can immediately see that the temperature dependence of the mass-gap
equation vanishes for any finite $T$ and we are left with the
zero-temperature mass-gap equation. On the other hand, for very high
temperatures and finite bare quark masses, there is only one solution, that is
$\vp_p=0$, which is the chiral restoration scenario.

An attractive feature of the GNJL model is
that the qualitative predictions of the model are robust against variations
of the quark kernel. We therefore consider, as a paradigmatic example, the
simplest quark kernel compatible with the requirements of confinement, 
\be
K_{\mu\nu}({\bm x}-{\bm y})=g_{\mu 0}g_{\nu 0}V_0({\bm x}-{\bm y}),\quad
V_0(r)=K_0^3r^2,
\label{simker}
\ee
and study the properties of the thermal mass-gap equation (\ref{mget0})
numerically. The harmonic oscillator confining potential
is known to lead to the simplest nontrivial mass-gap equation. Indeed, since the
Fourier transform of the 
quadratic potential (\ref{simker}) is given by the Laplacian of the 3D
delta-function, the mass-gap equation is differential 
\cite{Lisbon}. Switching from the
harmonic oscillator potential to another form of the potential amounts to simple
modifications of the quantitative
results, while all qualitative conclusions remain intact (see, for example,
Ref.~\cite{PedAl} for a detailed analysis).

Then the vacuum energy is given by the functional
\begin{eqnarray}
E_{\rm vac}&=&-\frac{g}{2}\int\frac{d^3p}{(2\pi)^3}
\biggl(A_p\sin\vp_p+B_p\cos\vp_p\biggr),
\label{Evacf}\\
A_p&=&m+\frac12C_F\int\frac{d^3k}{(2\pi)^3}V_0({\bm p}-{\bm k})\sin\vp_k,
\label{A}\\
B_p&=&p+\frac12C_F\int \frac{d^3k}{(2\pi)^3}\;
(\hat{{\bm p}}\hat{{\bm k}})V_0({\bm p}-{\bm k})\cos\vp_k.
\label{B}
\end{eqnarray}
Here $C_F=\frac12(N_C-1/N_C)$ is the $SU(N_C)_C$ Casimir operator in
the fundamental representation and the degeneracy factor $g$
counts the number of independent quark degrees of freedom,
$g=(2s+1)N_CN_f=12$.
Also the substitution $\sum_p\to V\int\frac{d^3p}{(2\pi)^3}$ was made for the
sake of convenience.
The mass-gap equation takes the form:
\be
[A_p\cos\vp_p-B_p\sin\vp_p]-(T/2)(\delta S/\delta\vp_p)=0,
\label{mge0}
\ee
with $\delta S/\delta\vp_p$ given in Eq.~(\ref{dmge}) or, in the explicit
form,
\begin{eqnarray}
K_0^3\left[\vp''_p+\frac{2\vp_p'}{p}+\frac{\sin
2\vp_p}{p^2}\right]=2p\sin\vp_p-2m\cos\vp_p\nonumber\\
-\frac{T}{2}\sin\vp_p\left[2\ln\left(\tan^2\frac{\vp_p}{2}\right)+\cos\vp_p\ln
2\right].
\label{mge1}
\end{eqnarray}
In what follows we consider the chiral limit and set the quark current mass $m=0$.

\begin{figure}[t]
\begin{center}
\epsfig{file=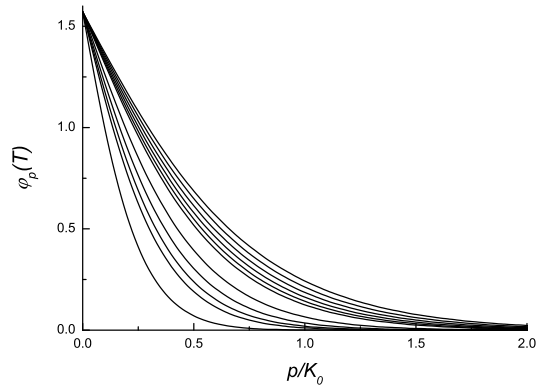,width=7cm} 
\caption{Solutions to the mass-gap equation (\ref{mge1}) for the temperatures $T=0$ (the upper curve), 
$0.2K_0$, $0.4K_0$, $0.6K_0$, $0.8K_0$, $K_0$, $2K_0$, $3K_0$, $4K_0$, $5K_0$, and $10K_0$ (the lower curve).}
\label{vpfig}
\end{center}
\end{figure}

\begin{figure}[t]
\begin{center}
\epsfig{file=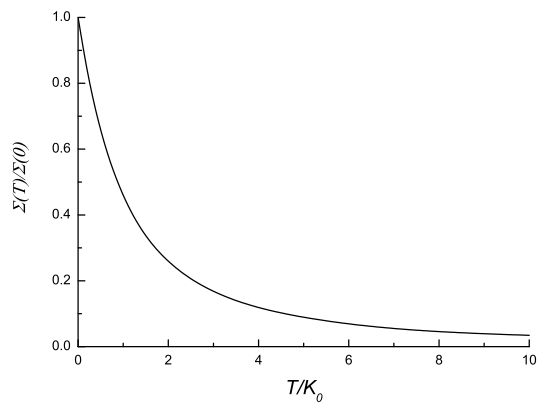,width=7cm} 
\caption{The chiral condensate $\varSigma(T)$ (normalised to the that at $T=0$) 
as a function of $T$ (measured in the units of the $K_0$).}
\label{condfig}
\end{center}
\end{figure}

In Fig.~\ref{vpfig} we plot solutions of Eq.~(\ref{mge1}) for various values
of the temperature. 
From this figure one can see that, as was anticipated before, with the increase of the temperature, the chiral angle
approaches chirally symmetric trivial solution. In addition, in Fig.~\ref{condfig}, we plot the temperature dependence 
of the chiral condensate
\be
\varSigma(T)=\langle\bar{q}q\rangle=-\frac{3}{\pi^2}\int_0^\infty dp\;
p^2\sin\vp_p(T),
\label{chcond}
\ee
which also demonstrates the same pattern of chiral symmetry restoration.

\section{Discussion}\label{sec4}

In this paper we studied the GNJL model at finite temperatures. The effect of
pair creation affected by both quark selfinteractions and temperature, is
conveniently described in the
formalism of the Bogoliubov--Valatin transformations, parametrised by the order
parameter: the thermal chiral angle
$\vp_p(T)$. At $T=0$ spontaneous breaking of chiral symmetry happens as a result
of such ${}^3P_0$ pairs condensation
in the vacuum. The resulting true (BCS) vacuum possesses a lower vacuum energy
than the trivial unbroken vacuum. At
finite temperatures, two processes proceed in opposite directions: confining
interaction tends to condense 
more ${}^3P_0$ quark--antiquark pairs in the vacuum, that is to increase the
chiral angle, while the effect of the 
temperature is a suppression of the latter. The balance of these two tendencies
is encoded in the thermal mass-gap 
equation (\ref{mget0}). Notice that, since the mass-gap equation is a nonlinear
equation, then the interplay of these 
two effects is essentially nonlinear. For low $T$'s the full thermal mass-gap
equation can be split, however, into two 
parts: the nonlinear mass-gap equation which describes chiral symmetry breaking
and the linear thermal equation which 
describes its partial restoration due to the temperature. The resulting chiral
angle is therefore just a sum of the 
zero-$T$ angle $\vp_p(T=0)$ and the thermal correction $\delta\vp_p(T)$. Let
us stress once more that such a 
disentanglement is only possible at $T\ll K_0$, where $K_0$ is the natural
energy scale introduced to the theory by 
the confining interaction. For high temperatures ($T\gg K_0$), chiral
symmetry
restoration is expected to take place, so the 
thermal mass-gap equation can be simplified under the assumption $\vp_p(T)\ll
1$. However, for $T\sim K_0$, the full nonlinear thermal mass-gap
equation has to be considered and 
solved.

As soon as the thermal mass-gap equation is solv\-ed and the chiral angle is
found, it can be used in
Bethe--Salpeter equations for hadrons or, alternatively, a second, generalised
Bogoliubov-like transformation can be performed in order to diagonalise, for the
hadronic sector, the Hamiltonian of the theory therefore enabling us to build
bound--state equations for mesons 
\cite{replica4}. The proposed real-time approach is general and can be applied
to any microscopic model for
confined systems. It allows one to investigate microscopically the proper
balance between chiral symmetry breaking due to confinement and chiral symmetry
restoration due to the temperature. 

A brief comment on the temperature behaviour of the model is in order. An
intrinsic
feature of the GNJL model, with confinement provided by an infinitely rising
potential, is that chiral symmetry breaking happens for any temperature $T$,
thus the chiral symmetry restoration process being asymptotical. This is related
to the fact that temperature effectively reduces the confining interaction
strength, while even quite weak confinement is able to break chiral symmetry.
Notice however that any realistic interquark interaction is expected to be
temperature-dependent. In particular, lattice calculations \cite{2} support the
conjecture \cite{3} that, while colour--magnetic vacuum fields do not change
across the deconfinement phase transition, the QCD vacuum loses its confining
colour--electric part. If this feature is incorporated into the quark kernel,
the resulting mass-gap equation will describe chiral phase transition.

The small-$T$ behaviour of the model also deserves special attention. From
general considerations (see Ref.~\cite{t1} for a review) one expects the
following behaviour of the chiral condensate at small $T$'s: 
\be
\Sigma(T)/\Sigma(0)=1-T^2/(8f_{\pi}^2)+\ldots.
\label{chc}
\ee 
However, since
$f_\pi\sim\sqrt{N_C}$, such corrections $O(T^2)$ appear only when the
theory is considered in the order $1/N_C$ or higher, that is beyond the
mean-field approximation --- see, for example, Ref.~\cite{t2}. 
In the GNJL model the mean-field approximation corresponds to the BCS level,
considered in this paper, while proceeding beyond the BCS approximation
corresponds to the full diagonalisation of the Hamiltonian (\ref{H}), including
the $:H_4:$ part --- see Eq.~(\ref{H3}). The details of this diagonalisation
procedure can be found in Ref.~\cite{replica4}. The correction to the
chiral condensate at small temperatures found in this paper is linear in $T$
which differs from the behaviour obtained in other models (see, for
example, the discussion of the NJL model in Ref.~\cite{t2}).
This should not
come as a surprise however since the BCS vacuum of the model behaves like an
infinite medium which is disturbed by the temperature, so that the response of
this medium, in the leading order, is linear in the perturbation, that is in
$T$. The medium has to possess rather peculiar features in order to respond
quadratically or in higher powers of the temperature, while we do not
observe such properties of the BCS vacuum. However, beyond the BCS
approximation, the GNJL model is subject to a severe realignment, so
that it finally acquires a description entirely in terms of mesonic states
\cite{replica4}, with the massless pions being the lowest states in the
spectrum. Then, with the free energy calculated
in terms of a pion gas, the standard approach can be applied to recover the usual
behaviour (\ref{chc}).
Given a complicated form of the Bogoliubov-like transformation which
diagonalises the Hamiltonian (\ref{H}) up to the order $1/N_C$ (to be confronted
to the $1/\sqrt{N_C}$ diagonalisation at the BCS level), the cancellation of the
linear in the temperature term in the chiral condensate does not look
unnatural. Indeed, in the mesonic vacuum, only properly correlated pairs
of mesons can be created or annihilated which appears to be a more involved
process than $^3P_0$ quark--antiquark pairs creation/annihilation in the BCS
vacuum. Thus the mesonic vacuum is expected to respond weaker to the temperature
increase, as compared to the quark vacuum.

Notice that the approach suggested in this paper is very different from other,
commonly adopted approaches to investigation of the thermal properties of QCD.
Indeed, while the ``standard'' procedure amounts to indirect tests of the
vacuum through the properties of the hadrons created on top of the vacuum, in our
approach the vacuum is probed directly. This makes important further  
investigations of the interplay between our real-time approach and the standard
imaginary-time approach to finite temperatures in confined systems.
We leave this problem for future publications. 
\bigskip

The authors would like to thank A. Abrikosov Jr., L. Glozman, Yu.
Simonov, and V. Vieira for
useful discussions. A. N. would like to thank the staff of
the Centro de F\'\i sica das Interac\c c\~oes Fundamentais
(CFIF-IST) for cordial hospitality during his stay in Lisbon,
where this work was originated and to acknowledge the 
support of the State Corporation of Russian Federation ``Rosatom'' as well as
of the grants RFFI-09-02-00629a, RFFI-09-02-91342-NNIOa, DFG-436 RUS 
113/991/0-1(R), NSh-843.2006.2, PTDC/FIS/70843/2006-Fi\-si\-ca, and of
the non-profit ``Dynasty'' foundation and ICFPM.

\end{document}